# First Experimental Demonstration of a 3-Dimensional Simplex Modulation Format Showing Improved OSNR Performance Compared to DP-BPSK

**Annika Dochhan[1], Helmut Griesser[2], Michael Eiselt[1]**
*(1) ADVA Optical Networking SE, Maerzenquelle 1-3, 98617 Meiningen, Germany,*
*(2) ADVA Optical Networking SE, Fraunhoferstr. 9a, 82152 Martinsried / Munich, Germany*
*adochhan@advaoptical.com*

**Abstract:** We experimentally demonstrate a novel 3-dimensional modulation format with 1.2-dB OSNR tolerance improvement potential compared to DP-BPSK, as verified for the linear case. The non-linearity tolerance is evaluated in single-span transmission over 300km SSMF.

## 1. Introduction

Transmission over very long distances requires not only a low noise transmission line with only small distortions but also a modulation format, which exhibits a high tolerance to optical noise. Recently, 4-dimensional modulation formats have been proposed [1], in which both signal polarizations are co-modulated, resulting in an improved OSNR tolerance. In this paper, we report on a novel modulation format exhibiting the same spectral bandwidth as dual-polarization (DP-) BPSK, but with an improved OSNR tolerance. Signals with this modulation format are used to transmit over 300 km of standard-single mode fiber with an insertion loss of 63 dB, using only counter-directional Raman pumping and EDFA amplification.

## 2. Novel 4-dimensional modulation format

An optical carrier wave can be modulated in four orthogonal dimensions: I- and Q- phases of two orthogonal polarizations. In the standard DP-BPSK or DP-QPSK modulation formats, each of these dimensions is independently modulated with a binary signal, resulting in the transmission of one bit per dimension or 2 or 4 bits per symbol. In a 3- or 4-dimensional format, the modulation of the orthogonal dimensions is interdependent. Tab. 1 shows the constellation coordinates of the novel 3D-Simplex format. Main advantage of this representation is that it allows unbiased binary driving signals (see Figure 1). It can be seen that two bits are encoded in a QPSK format in the x-polarization, while the y-polarization is BPSK modulated with the XOR of two input bits. The minimum Cartesian distance between two points is $D_{min}$=sqrt(8) with an average symbol power of $P_{avg}$=3. Therefore, in the limit of low bit error rates, an improvement of the OSNR tolerance over DP-BPSK ($D_{min}$=sqrt(4), $P_{avg}$=2) of 1.2 dB is expected.

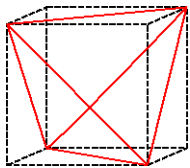

Fig. 1: 3-D Simplex in a cube, corners representing the constellation points in three dimensions (4th dimension is not used).

| Bit encoding | Ix | Qx | Iy | Qy |
|---|---|---|---|---|
| 00 | -1 | -1 | -1 | 0 |
| 01 | -1 | 1 | 1 | 0 |
| 10 | 1 | -1 | 1 | 0 |
| 11 | 1 | 1 | -1 | 0 |

Tab. 1: Constellation points of 3D-Simplex modulation format

## 3. Experimental setup

The novel modulation format was first tested in a back-to-back configuration for 16 and 25 GBaud, the latter being a possible line rate for 40 Gb/s long-haul transmission with sufficient FEC overhead. In addition, at 16 GBaud, single span transmission over 300 km, evaluating the non-linearity tolerance of the format has been performed. All results are compared to DP-BPSK with differential encoding at the same symbol rate. In contrast, the Simplex format is not invariant for certain phase rotations and therefore does not require differential encoding. The experimental setup is shown in Figure 2. Driving signals for a DP-IQ Mach-Zehnder modulator were generated offline and stored in a DAC (64 GS/s, 13 GHz bandwidth) [2]. The x-polarization was QPSK modulated with a $2^{11}$ de Bruijn sequence, while the I-branch of the y-polarization was BPSK modulated with the XOR combination of the two QPSK encoded bits. The Q-branch of the y-polarization was left unmodulated and the bias was set to blocking. Due to the limited bandwidth of the used DAC and the real time sampling scope on the receiver side, initial experiments were performed with a symbol rate of 16 GBaud. The back-to-back performance of 25-GBaud signals was also tested, but shows an increased implementation penalty. The transmission link consisted of 300 km with an attenuator located after 100 km. The fiber launch power was varied between 10 and 20 dBm. Counter-directional 4-wavelength Raman pump power of 29 dBm was injected from the receiving end. At the receiver, the signal was filtered by a 35-GHz (at 16 GBaud) or 65-GHz (at 25 GBaud) optical bandpass filter (Finisar WaveShaper), emulating the demultiplexer in 50-GHz or 100-GHz WDM transmission. The signal was received using a standard coherent receiver, comprising a local oscillator laser, a polarization beam splitter, two 90° hybrids and four balanced photo diodes. The linewidths of both, transmit and receive lasers, were approximately 100 kHz. The offline receiver is based on a purely blind equalization algorithm, with the initial polarization ambiguity assumed to be solved by a higher level detection mechanism or a dedicated training sequence. Therefore, in the experiment we had to rely on a coarse manual polarization alignment. For DP-BPSK, the coefficients of the butterfly structure FIR filter were updated using the observation of two consecutive output symbols as described in [3]. The 3D-Simplex format required a combination of the BPSK-algorithm from [3] and the standard CMA [4]. Dependent on the polarization rotation, the phase between the BPSK and the QPSK tributary has to be estimated carefully. Other components of the offline DSP include standard algorithms for dispersion compensation, clock recovery, frequency offset compensation, and phase estimation.

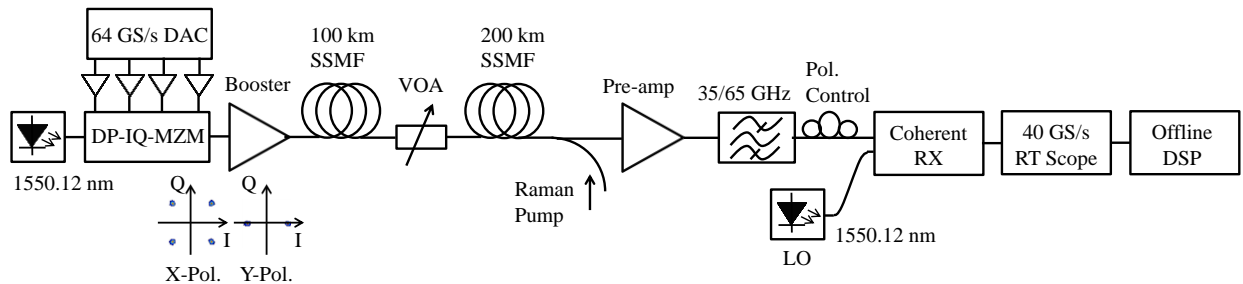

Fig. 2: Experimental setup. A DP-QPSK Mach-Zehnder-Modulator is used to generate a QPSK signal in the x-polarization and a BPSK signal in the y-polarization. The signal is transmitted over a single 300 km span of SSMF using backward Raman pumping.

### 4. Results

Received constellations of the 3D-Simplex format at an OSNR of 15.9 dB and back-to back results for DP-BPSK and 3D-Simplex are shown in Figure 3 (solid lines, markers) together with theoretically expected values (dashed lines). From figure (c) it can be seen that the implementation penalty at 16 GBaud for both formats is about 1 dB and that the use of 3D-Simplex reduces the OSNR requirements by about 1 dB compared to DP-BPSK. This benefit is maintained at 25 GBaud (d), while the implementation penalty increases due to the transmitter bandwidth limitations mentioned above. The OSNR is defined with a 0.1 nm noise bandwidth, the signal power was measured in a bandwidth of 0.5 nm.

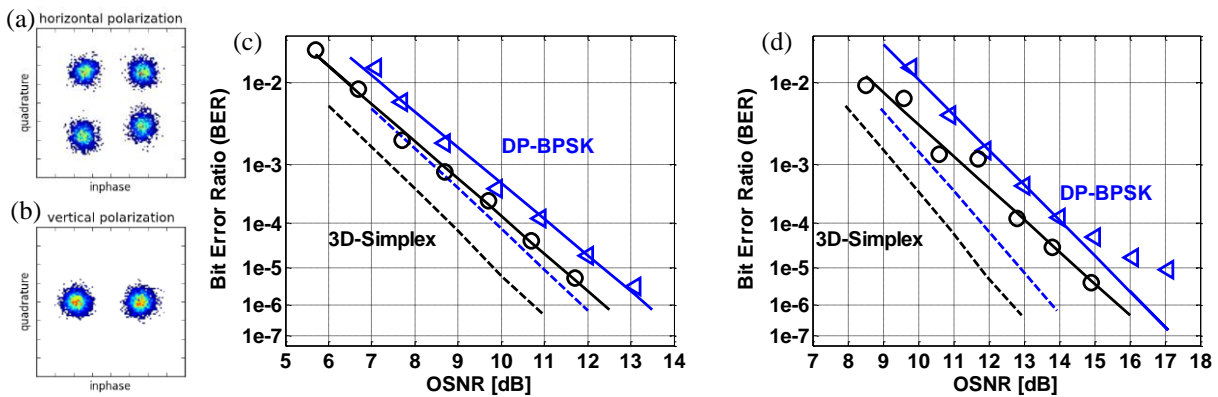

Fig. 3: Received Simplex constellations for an OSNR of 15.9 dB after DSP for (a) QPSK tributary, (b) BPSK tributary, and back-to-back BER results for (c) 16 GBaud and (d) 25 GBaud. 3D-Simplex (circles) in comparison to DP-BPSK (triangles). Dashed lines: theoretical values, markers: experimental results (solid lines: linear regression).

The 16 GBaud signal was then transmitted over 300 km of standard single-mode fiber with varying input power. Figure 4 (left) shows the resulting bit-error ratio (BER) of the 3D-Simplex and the DP-BPSK signal as a function of the fiber launch power. Without additional attenuation, the loss of the 300 km of fiber was approximately 63 dB. The optimum input power for DP-BPSK turned out to be 17 dBm while 3D-Simplex was optimized for 16 dBm input power. Figure 4 (right) evaluates the BER at optimum launch power as a function of the fiber loss (which can be increased using the variable optical attenuator after 100 km of fiber) for both modulation formats. Here, for DP-BPSK the lowest shown attenuator loss of 0 dB corresponds to an OSNR of 13.9 dB, while the OSNR of 9.9 dB is equivalent to an additional loss of 4 dB. For 3D-Simplex 0 dB attenuator loss leads to an OSNR of 12.9 dB, and 9.9 dB OSNR is equivalent to 3 dB attenuator loss.

It can be seen that the 3D-Simplex modulation format exhibits 1 dB less nonlinear tolerance as compared to DP-BPSK. The measurements for various fiber loss values show that the benefit which was seen back-to-back is not maintained. This can be explained by the sensitivity of 3D-Simplex towards phase distortions, which also deteriorates the performance of the blind equalizer. It is expected that a training sequence based equalization has the potential to improve the non-linearity tolerance of the 3D-Simplex format, so that the benefit, which is apparent in the back-to-back results, can be maintained after transmission in the non-linear regime. However, the non-linear tolerance of DP-QPSK (which only adds two more constellation points) would be even several dB worse.

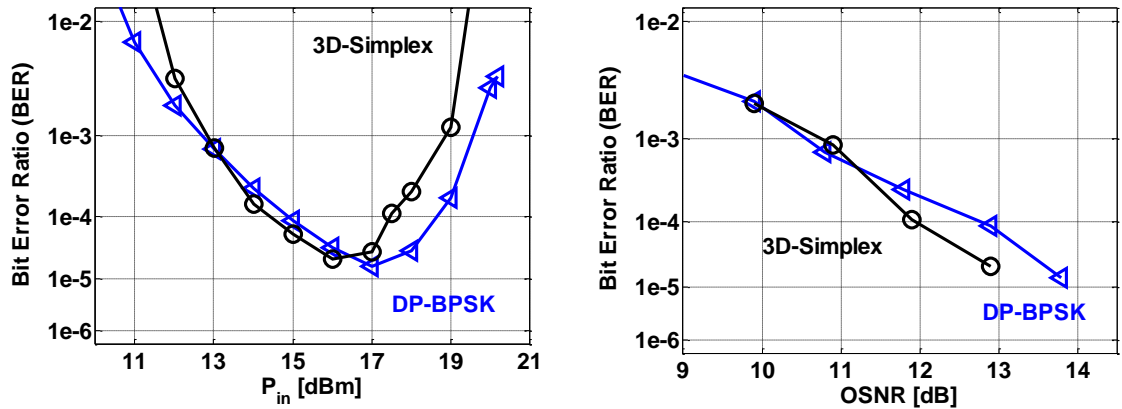

Fig. 4: Transmission of 3D-Simplex (circles) in comparison to DP-BPSK (triangles). Left: BER vs. fiber input power $P_{in}$. Right: BER vs. OSNR with 300 km transmission at optimum launch powers (DP-BPSK: 17 dBm, 3D-Simplex: 16 dBm), increasing span loss (and thus reducing OSNR). For DP-BPSK an OSNR of 13.9 dB corresponds to 0 dB additional loss, for 3D-Simplex OSNR of 12.9 dB is equivalent to 0 dB loss.

## 5. Conclusion

We have demonstrated the improvement of the OSNR tolerance of 2-bit per symbol transmission by 1 dB using a novel 3D-modulation format. Transmission over a long span with a high launch power demonstrated that the resilience against non-linear effects is comparable to that of DP-BPSK. To our knowledge, this is the first demonstration of a 3D-modulation format encoding 2 bits per symbol and improving over the OSNR tolerance of DP-BPSK. This novel modulation format is therefore a promising solution for ultra-long haul and submarine transmission systems, where the OSNR tolerance is crucial.

## 6. Acknowledgments

We thank our colleague Joerg-Peter Elbers for fruitful discussions. The research leading to these results has been funded in part by the German ministry for education and research (BMBF) under Grant 01BP12400 as part of the ADVAntage-NET project.